# *'Static' Optics*

*Ahmed M. Mahmoud, Nader Engheta*[*]

*Department of Electrical and Systems Engineering*
*University of Pennsylvania*
*Philadelphia, Pennsylvania 19104, USA*

**Abstract**

**In recent years, the concept of metamaterials has offered platforms for unconventional tailoring and manipulation of the light-matter interaction. Here, we explore the notion of "static optics", in which the electricity and magnetism are decoupled and their fields are statically distributed, while being temporally dynamic. This occurs when both the relative effective permittivity and permeability attain near-zero values at an operating frequency. We theoretically investigate some of the unprecedented wave features, such as unusual radiation characteristics of an emitter embedded in such epsilon-and-mu-near-zero (EMNZ) media. Using such static-optical medium one might in principle 'open up' and 'stretch' the space, and have regions behaving electromagnetically as 'single points' despite being electrically large. We suggest a possible design for implementation of such structures using a single dielectric rod inserted in a waveguide operating near its cut-off frequency, providing the possibility of having electrically large "empty" volumes to behave as EMNZ static-optical media.**

Throughout the last decade, there have been extensive research efforts in the field of metamaterials [1-6] – artificial structures that exhibit unusual properties that do not readily exist in nature. Recently there has been great interest in the near-zero parameters materials. This is the category of metamaterials whose relative permittivity is near zero (i.e., epsilon-near-zero (ENZ)), or relative permeability is near zero (i.e., mu-near-zero (MNZ)), or both relative permittivity and permeability are near zero (i.e., epsilon-and-mu-near-zero (EMNZ)). The special feature of such media basically lies in having a low wave number (i.e., stretched wavelength) as a consequence of the near-zero refractive index that leads to a relatively small phase variation over physically large region of such media. This opens the door to various interesting wave phenomena and applications. One track in which this category of metamaterials was utilized efficiently is the field of antenna design, where ENZ or EMNZ materials were considered for tailoring the radiation patterns, i.e., to attain highly directive radiation patterns [7-13], or for significantly enhancing the radiation efficiency [14-16]. On the other hand, near-zero-parameters materials have also been extensively studied and used as means to realize unconventional tunneling of electromagnetic energy within ultra-thin sub-wavelength ENZ channels or bends (a phenomenon coined as supercoupling) [17-20], tunneling through large volumes using MNZ structures [21], and to overcome the problem of weak coupling between different electromagnetic components that are conventionally not well matched, e.g. in a coaxial to waveguide transition [22]. Another interesting field of research has been about manipulating the transmission characteristics using near zero media [23-26] where such media have been loaded by certain properly designed inclusions that allow shaping of the system's transmission profile. Recently, the field of anisotropic near-zero parameters media has been explored, adding extra degrees of freedom and thus achieving some interesting phenomena [27-30]. Moreover, zero parameters materials have been integrated with non-linear elements to achieve controllable transmission [31] or to enhance the non-linear response of Kerr-based structures in the optical regime [32, 33]. Owing to the numerous potential applications and novel physical phenomena that can result from such media, great efforts have been put into the actual realization of those media, from naturally available materials [34-36] to using photonic crystals that exhibit dirac cones dispersion [37, 38] and structures in the microwave regime [39, 40], and in the more challenging optical regime [41] where gain media has usually been exploited to overcome the problem of high losses [42, 43].

---

[*] To whom correspnding should be addressed, email: engheta@ee.upenn.edu

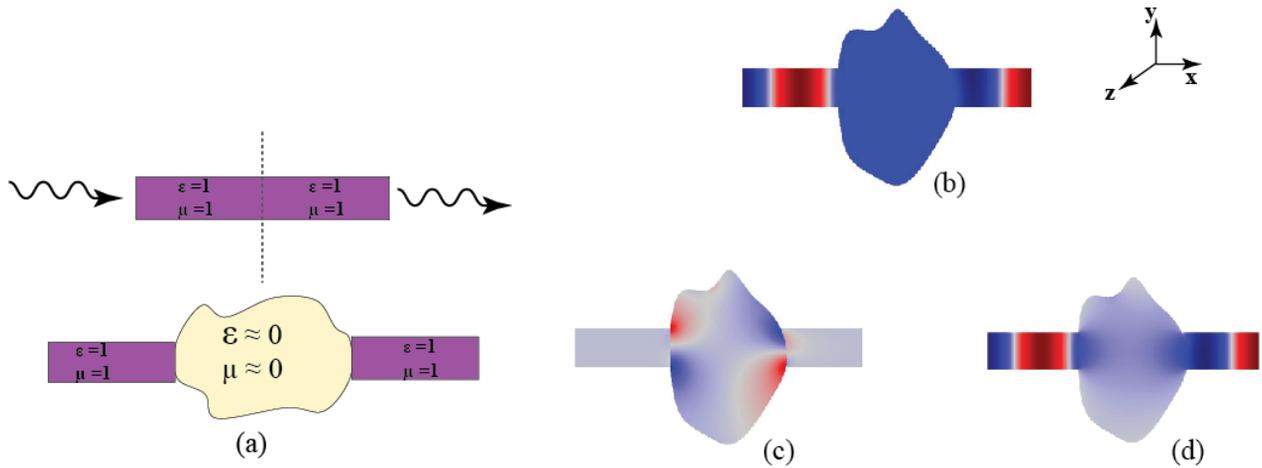

*Fig. 1: EMNZ medium as an "electromagnetic point":* *(a) Schematic of an epsilon-and-mu-near-zero (EMNZ) medium emulating opening up or stretching the space at the dotted line between the two sections of an air-filled parallel-plate waveguide, while keeping the region as a "single point" electromagnetically. (b) Snapshot of the z-component of the magnetic field distribution (shown with blue color) and the x-component (c) and y-component (d) of the electric field distributions over the proposed structure with an arbitrarily shaped cross section filled with an idealized EMNZ material. The external walls (except at the exits of the two ports) are made of perfectly electric conductor (PEC).*

In the present work, we shed light on some of the exotic phenomena of EMNZ media, providing a study regarding two of the main pillars of electromagnetic behavior within EMNZ environments, namely its scattering properties and its interaction with emitting dipoles. Moreover, building on the findings in earlier works [18] we propose a design scheme that provides us with both effective permittivity and effective permeability near zero over a relatively large "empty" volume, and discuss the limitations. Since in such an EMNZ medium, both $\nabla \times E = 0$ and $\nabla \times H = 0$ simultaneously, the electric and magnetic phenomena are decoupled and spatially distributed statically, while still temporally dynamic. This leads to a paradigm in which we may have optical phenomena while the field distributions are static-like, effectively having a scenario as "DC optical circuits". Having such an EMNZ region, as shown in Fig.1(a), we can break an air-filled parallel-plate waveguide at a point, and effectively open up or stretch the space between two parts of this waveguide, without affecting external electromagnetic entities and quantities, implying that we can have an electromagnetically large physical volume, that would have otherwise influenced the electromagnetic wave propagation externally, but now owing to the EMNZ effect it behaves as if it is a 'single point' electromagnetically as viewed from the external world. The external boundary of the entire region shown in Fig. 1 (except for the input and output exit ports) is PEC wall. As depicted in a two-dimensional (2D) simulation shown in Fig. 1(b), perfect transmission (with unit magnitude and zero phase difference) from the input to the output port is preserved regardless of the arbitrary shape and size of the EMNZ region. Moreover, in this 2D simulation of the EMNZ region, the magnetic field spatial distribution is uniform while the electric field is distributed spatially as though a "battery" is connected between the top and the bottom plates of the EMNZ region.

An interesting phenomenon that takes place within EMNZ environments is the unconventional scattering performance of perfectly electric conducting (PEC) objects embedded in an EMNZ medium. Recently, there has been some works investigating the behavior of zero-index media when loaded by various structures and the ability of these systems to manipulate the transmission and reflection profile owing to the unique scattering performance within zero-index media [23-26] . Here, using the two-dimensial Mie scattering theory we study the unique scattering performance from two-dimensioanl PEC objects, embedded in a 2D host EMNZ medium. We begin by analytically solving for the scattered fields from a 2D PEC cylinder embedded in an unbounded EMNZ medium for both the transverse magnetic (TM, i.e., the incident H vector is parallel with the cylinder axis) and transverse electric (TE, i.e., the incident E vector is parallel with the cylinder axis) incident waves normally incident on the cylinder. For a TM wave, the incident and the scattered fields can be written (assuming $e^{-i\omega t}$ and that the axis of the cylinder is along the z axis) in the time harmonic case as follows [44]

$$\vec{H}^i = \hat{z} H_o \sum_{n=-\infty}^{\infty} i^n J_n(\text{k}_{EMNZ}\text{r}) e^{in\phi} \quad (1)$$

$$\vec{H}^s = \hat{z} H_o \sum_{n=-\infty}^{\infty} c_n^{TM} H_n^{(2)}(\text{k}_{EMNZ}\text{r}) \quad (2)$$

Applying the boundary conditions we find that

$$c_n^{TM} = -i^n \frac{J_n^{'}(\text{k}_{EMNZ}\text{a})}{H_n^{'(2)}(\text{k}_{EMNZ}\text{a})} e^{in\phi} \quad (3)$$

Similarly, for the TE case, we can find that

$$c_n^{TE} = -i^n \frac{J_n(\text{k}_{EMNZ}\text{a})}{H_n^{(2)}(\text{k}_{EMNZ}\text{a})} e^{in\phi} \quad (4)$$

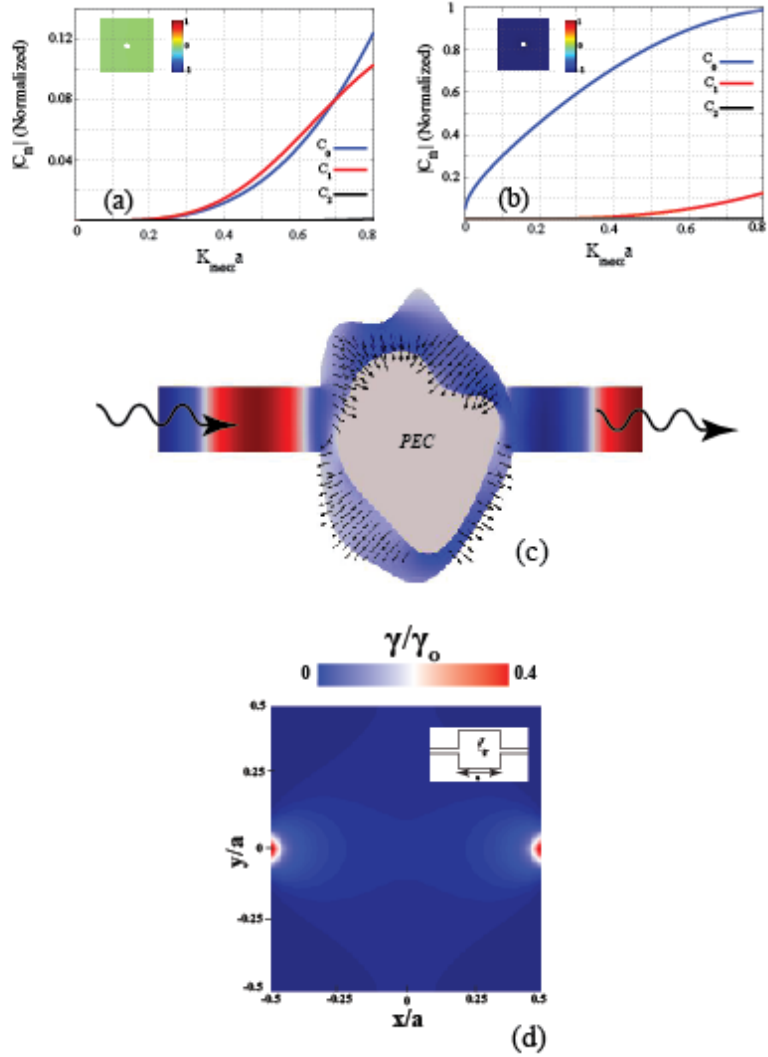

*Fig. 2: Mie scattering coefficients $|C_n|$ vs $k_{EMNZ}$ around the strict condition of EMNZ for (a) TM mode, (b) TE mode. Two-dimensional scattered magnetic field from a 2D conducting cylinder of radius $5\lambda$ embedded in an ideal EMNZ medium when illuminated by TM mode (inset (a), and two-dimensional scattered electric field from a conducting cylinder of radius $5\lambda$ embedded in an ideal EMNZ medium when illuminated by TE mode (inset (b)). (c) Snapshot of the two-dimensional electric (shown as arrows) and magnetic (shown as blue color) field distributions in an EMNZ region, loaded with a conducting cylinder, both with arbitrarily shaped cross sections. External walls (other than the exits of the two ports) are assumed to be PEC walls. (d) Distribution of normalized spontaneous emission rate of a two-dimensional electric dipole inside the EMNZ enclosure with two ports connected to parallel-plate waveguides filled with air, versus the dipole position within the structure. The geometry of the EMNZ enclosure and the two air-filled parallel-plate waveguides are shown in the inset.*

We investigate the dependence of $|C_n|$ on $k_{EMNZ}$ as depicted in Fig.2(a) and (b). As shown, there exists an interesting difference between the scattering behavior of the cylinder being illuminated by a TM wave versus a TE wave. For the TM mode as shown in Fig. 2(a) the dependence of $|C_n|$ on $k_{EMNZ}$ is relatively weak around the strict zero condition, meaning that for the case where $k_{EMNZ}$ is small but not strictly zero we can still consider having zero $|C_n|$ for all *n*'s. Thus for that mode a conducting cylinder would introduce no scattering and would be totally cloaked. (An intuitive way to appreciate this point is to consider the fact that in the EMNZ medium, $D=0$, and thus $\nabla \cdot D = 0$. Therefore, no charge can be induced on the surface of this 2D PEC cylinder when illuminated by the TM wave.). On the other hand, Fig. 2(b) shows that for the TE mode $|C_0|$ exhibits a relatively stronger dependence on $k_{EMNZ}$ in vicinity of the strict zero condition, suggesting that for the case where $k_{EMNZ}$ is near zero, but not strictly zero, we should expect a strictly non-zero scattering from the cylinder with the scattered fields being azimuthally symmetric, when illuminated by that mode. These findings are verified numerically in the insets of Fig. 2(a) and (b) where in (a) we see the scattered fields from a PEC cylinder of radius 5λ, with λ being the free space wavelength, embedded in an EMNZ medium, and illuminated by a TM wave. As expected the scattered fields due to the 2D PEC cylinder vanish. On the other hand, the inset of Fig. 2(b) shows the scattered fields from a conducting cylinder of the same radius that is illuminated by a TE wave. It is shown that for that polarization we get a non-zero scattering only for the *n*=0 component of the scattered field when the relative permittivity and permeability are near zero (but not strictly zero), as expected. Moreover, the interesting property of PEC objects introducing no scattering being illuminated by a TM wave when embedded in an EMNZ medium still holds regardless of how arbitrarily shaped the EMNZ region and the conducting object are, owing to the ability of EMNZ media to open up the space as discussed before. This is depicted clearly in Fig. 2(c), where a physically large-cross-section 2D PEC cylinder of an arbitrary shape is embedded into the EMNZ region with another arbitrary shape connected to two air-filled parallel-plate waveguides, shown in Fig. 1(b), providing no scattering at all, with unity transmission and almost no phase progression as the wave traverses the EMNZ region containing this cylinder. (As in Fig. 1, the external boundary (except for the input and output ports) is made of PEC wall. The two parallel-plate waveguides connected to the 2D EMNZ region are filled with air).

    In this section we investigate the behavior of emitting dipoles within arbitrarily shaped EMNZ media, and more specifically how the EMNZ environment influences the spontaneous emission rate of emitters leading to some interesting unprecedented behavior. In [17] the phenomenon of supercoupling that relies on energy squeezing through subwavelength narrow channels has been introduced in epsilon-near-zero (ENZ) metamaterials. Owing to the substantial electric field enhancement within the narrow ENZ

channel, placing an electric-dipole emitter within that ENZ channel leads to significant emission enhancement that can be quantitatively described in terms of the dipole's spontaneous emission rate $\gamma$ as compared to its emission rate in free space $\gamma_o$ i.e., $\gamma/\gamma_o$ [45, 46]. However, a unique feature of such structure as compared to any other resonant system coupled to a radiating element or molecule is the nearly uniform phase distribution of that enhanced electric field along the channel, which is in principle independent of the channel length. This leads to the independence of the emission enhancement on the dipole's location within the channel. In an EMNZ medium on the other hand, the supercoupling phenomenon is still preserved, however, without the need for ultranarrow subwavelength channels as was the case for ENZ media, which in addition to providing a larger platform to place the dipole within the medium, leads to no mandatory enhancement in the fields within the EMNZ region. This may lead to the novel effect of emission inhibition regardless of the position of the dipole within the EMNZ region. To numerically demonstrate that phenomenon, we investigate the 2D structure shown in the inset of Fig. 2(d), where a two-dimensional y-oriented dipole is placed within a $5\lambda \times 5\lambda$ enclosure filled with an ideal EMNZ material connected to two narrow parallel-plate waveguides filled with air. (The cross-sectional shape of this 2D EMNZ region is chosen arbitrarily. The external boundary, except for the two ports, is make of PEC wall). As shown in Fig. 2(d) where a color map of two-dimensional distribution of $\gamma/\gamma_o$ versus the dipole location within the EMNZ region is shown, we get significant emission inhibition (i.e, $\gamma/\gamma_o$ less than unity) within the EMNZ region as expected. Owing to the reciprocity, this $\gamma/\gamma_o$ distibution resembles the distribution of the y-component of the electic field when this structure is fed with the transverse electromagnetic (TEM) mode incident from one of the air-filled parallel-plate waveguide port.

Now we explore a scenario in which a properly selected single dielectric inclusion within a host medium whose permittivity (or effective permittivity) is near zero (e.g., a large waveguide section with the $TE_{10}$ mode near its cut-off frequency) [47] may lead to simultaneously having effective permittivity and permeability near zero. We also explore the limits to which such structure could mimic the interesting characteristics discussed in the previous sections, including the ability to open up the space, the unique scattering performance of electrically large PEC structures embedded in it, and non-conventional radiation performance for radiating elements placed within the proposed structure. It was shown in previous work [18] that with an ENZ host medium one can also achieve an effective MNZ by periodically loading this host medium with inclusions of proper dimensions and permittivity. In this section we show that this can be extended into a non-periodic case with arbitrary shape of the cross section, and that even within one unit cell, which can be arbitrarily large and not limited to a sub-wavelength size, we can still have both effective permittivity and permeability near zero. We propose a practical design in which the interesting features of EMNZ can be studied in the microwave regime. At such frequencies, ENZ materials are not readily available in nature, however it has been shown in [48] that parallel metallic plates can simulate a two-dimensional artificial plasma when the $TE_{10}$ mode is considered. The effective permittivity of the such waveguide structures follows a Drude-like model, i.e., $\varepsilon_h/\varepsilon_0 = \varepsilon_d - (\frac{\pi}{k_0 d})^2$, where $\varepsilon_d$ is the relative permittivity of the dielectric between the metallic plates, $d$ is the separation between the metallic plates, and $k_0$ is the wave number in free space. In such an environment, if $d$ is chosen to be $\lambda_0/2$, where $\lambda_0$ is the free space wavelength at some chosen operating frequency $\omega_c$ we need to load this waveguide with a material of permittivity $\varepsilon_i + 1$, in order to emulate a material with relative permittivity $\varepsilon_i$ Thus as shown in Fig. 3(a) the input and output channels that are required to emulate an air-filled parallel plate waveguide with the TEM mode are mimicked using $TE_{10}$ mode in rectangular waveguides filled with a dielectric of relative permittivity of 2 (regions 1 and 3). In the proposed EMNZ region (region 2 in the middle) the ENZ host medium is emulated by a waveguide filled with dielectric of relative permittivity of 1. We choose the inclusion to be a circular-cylinder dielectric rod of radius R and permittivity $\varepsilon_i$, which when embedded periodically in an ENZ host, one can achieve an effective permeability shown in [18] to be

$$\mu_{eff} = \mu_0 (\frac{A_{h,cell}}{A_{cell}} + \frac{2\pi R^2}{A_{cell}} \frac{1}{k_i R} \frac{J_1(k_i R)}{J_o(k_i R)}) \qquad (5)$$

where $A_{cell}$ is the total area ($a \times a$) of the unit cell, $A_{h,cell} = A_{cell} - \pi R^2$, and $k_i = \omega\sqrt{\varepsilon_i \mu_0}$. (Note that this rod can be placed almost anywhere in the unit cell. Moreover, although in Fig. 3(a) the cross section of the middle section is taken to be rectangular, any arbitrary shape with the required area can be chosen.) As mentioned before, we are interested in a near zero effective permeability which for $a = 1.3\lambda$ with a dielectric rod of radius $0.13\lambda$ requires the rod's relative permittivity to be 9.35. We show that even using only one unit cell we still get almost perfect transmission at the operating frequency $\omega_c$ as numerically shown in Fig. 3(b) as opposed to getting almost no transmission in absence of the dielectric rod. Fig. 3(c) shows a cross section of the distribution of the magnitude of the magnetic field within the structure, and the almost perfect transmission between the two ports is evident here, with almost uniform phase within the proposed EMNZ region as shown in Fig. 3(d). Thus, it is numerically verified that our proposed structure acquires the desired EMNZ behavior of opening up the space with near unity magnitude and zero phase transmission between the two ports.

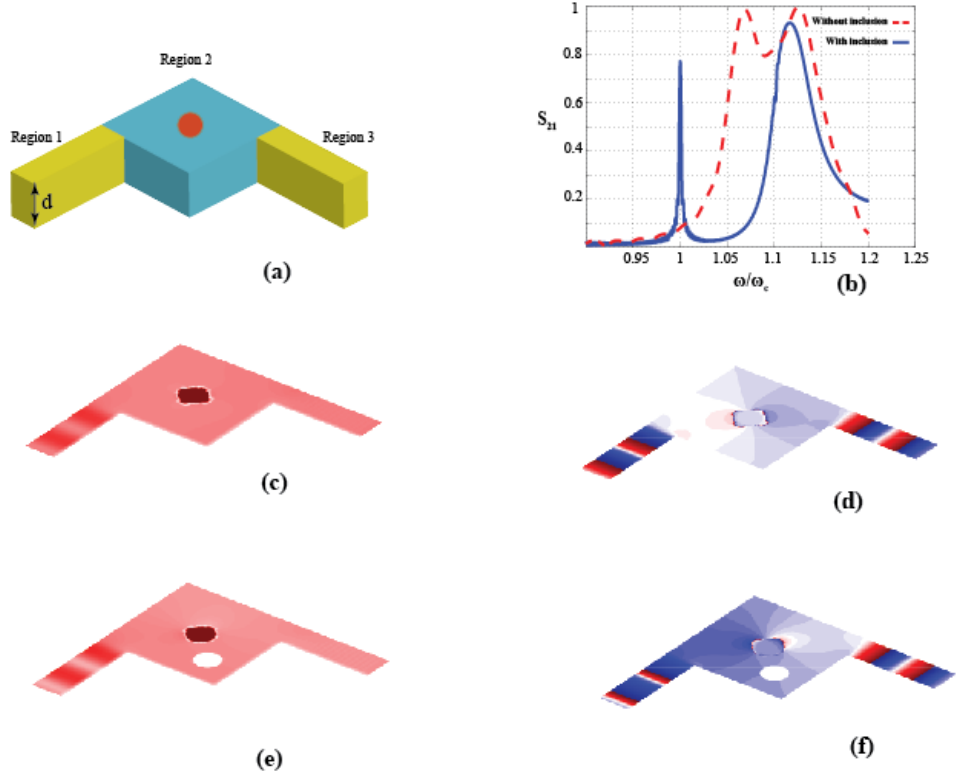

*Fig. 3: Proposed design for a structure imitating effectively an EMNZ medium: (a) Geometry of rectangular air-filled metallic waveguide structure (middle region, blue) with a dielectric rod inclusion (red) emulating the EMNZ behavior. This middle section is connected to two idential dielectric-filled metallic rectangular waveguides (yellow) as the two ports (b) Magnitude of the transmission coefficient ( $S_{21}$ ) of the structure shown in (a) vs normalized frequency, when fed by one of the waveguide ports. Here $\omega_c$ is the cut-off frequency of the waveguide section. Blue and red curves are for the cases of with and without the dielectric rod, respectively. (c) Amplitude and (d) phase of the magnetic field in the middle plane within the structure, when it is illuminated by a wave coming from one of the ports; (e) and (f) Amplitude and phase of the magnetic field in the middle plane within the structure when the PEC scattering cylinder is also inserted.*

Next, we proceed to investigate whether having a PEC cylinder embedded in our proposed effectively EMNZ structure would introduce no scattering for the incident mode with the magnetic field vector parallel with the axis of the cylinder as discussed previously. This is depicted clearly in Figs. 3(e) and (f) where a PEC cylinder of radius $0.13\lambda$ is embedded into the proposed structure. It is worth mentioning here that now loading the EMNZ unit cell with a conducting cylinder, the effective area of the unit cell is reduced, and that consequently for the same radius of a dielectric rod, the rod's permittivity has to be adjusted to take that into account. As depicted clearly in Fig. 3(e) and (f) the conducting rod does not introduce any scattering as expected.

Finally we investigate an electric-dipole emission performance within our proposed structure. The phenomena of opening up the space and the unique scattering performance within an EMNZ medium are effectively "global" phenomena, i.e., they can be described using effective media approaches. On the contrary, the dipole spontaneous emission rate is a phenomenon that is highly dependent on its surrounding environment. Thus one might expect the spontaneous emission inhibition phenomenon that was shown previously within a idealized EMNZ medium would not necessarily take place within our proposed structure, whose "effective" permittivity and permeability exhibit near zero values (while the local material parameters may differ from zero), but the radiation from the ports may be influenced by such effective EMNZ characteristics. We numerically examine the radiation of a given short electric x-oriented dipole placed in various locations inside our proposed design for the effectively EMNZ structure with two ports to the outside region. As shown in Fig. 4(a), where $\gamma/\gamma_o$ distribution as a function of the dipole's location within the proposed structure is given, there exists a significant enhancement in the spontaneous emission rate of such a dipole when it is located in proper locations within the structure. This enhancement is more pronounced towards the center of the structure, completely consistent with reciprocal scenario when one would get high values of electric field near the dielectroc rod, if the incident energy were fed as a $TE_{10}$ mode through one of the waveguide ports. This can also be explained by noting that as the dipole gets closer to the center, the fundamental mode of the dielectric rod (that is the mode for which effective zero permeability is achieved) is better coupled to the dipole emission. One of the highlights of the dipole emission in such an effectively EMNZ structure is the intriguing phenomenon of the equal distribution of the radiated power between the two output ports (due to the uniformity of the phase within the structure), regardless of the relative location of the dipole to either of the ports and regardless of the location of the ports connected to the structures. This is shown in Fig. 4(b), (c), and (d) where moving the dipole from the center to the corner or towards the left port does not introduce any change on the ratio between the powers exiting the two ports. Specifically, while the total radiated power from the dipole may depend on its location inside this structure (e.g., this total power is weaker in Fig. 4(c) and 4(d) than in Fig. 4(b)), this power is divided equally as it exits from the two ports when the two ports' cross sections are the same. This intriguing power division is independent of how close the dipole may be located near one of the ports and is also independent of where the two ports are conneted to the structure. As can be seen in Fig. 4(c), the waves exiting from the two ports are the same in each case (with the same amplitude and phase), even though the dipole is very close to the left port and far from the right port. This implies that the radiation from the dipole couples to the two ports the same way despite of unequal distance of the dipole from the two ports. This phenomenon may have important implications in quantum optics and the photon entanglement in such platforms. We will report on this issue in a future publication.

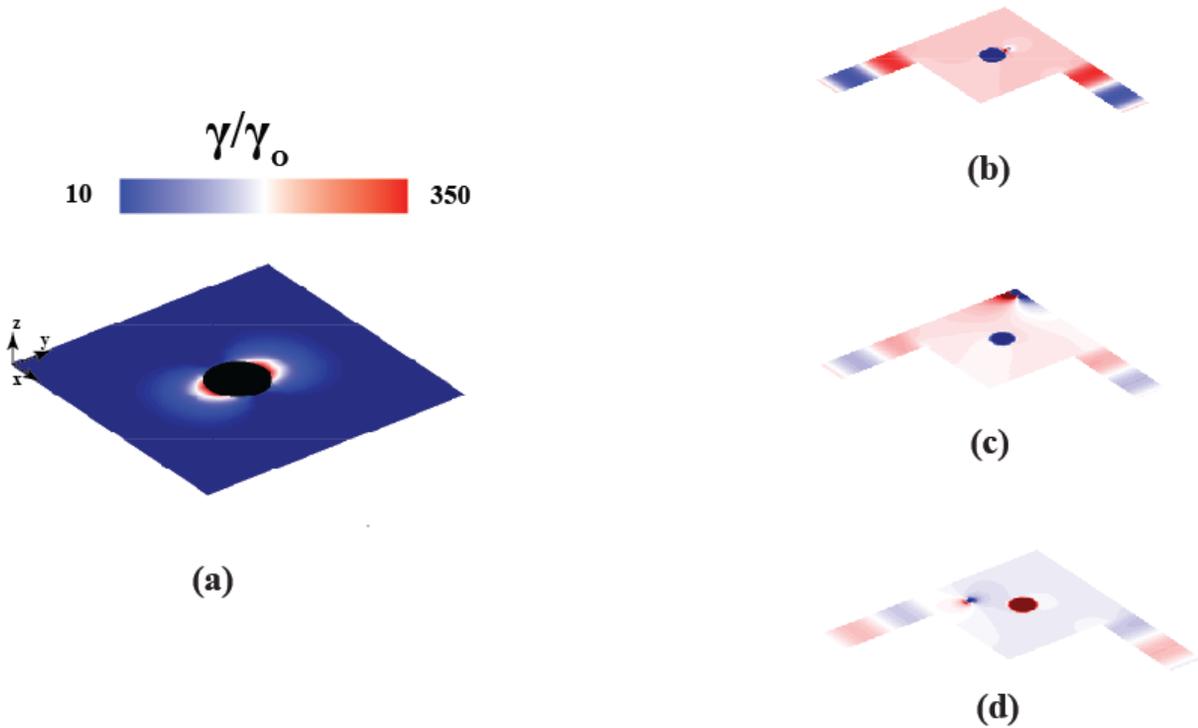

*Fig. 4*: ***Emission of a dipole located inside the proposed structure***: *(a) Distribution of the normalized spontaneous emission rate of the short electric dipole versus the dipole position within the designed enclosure imitating EMNZ medium. (b) A snapshot of the magnetic field of the dipole radiation within the structure with the short electric dipole placed close to the center. (c) as in (b) but the dipole is brough to the corner of the structure. (d) as in (b) but the dipole is brought close to the left port. While the total radiated power is different in (b) (c), and (d), the power exited from the two ports remains equal in each case, regardless of the location of the dipole within the structure and the relative location of dipole to the ports.*

In conclusion, we have explored some of the intriguing phenomena of dipole emission and wave scattering in EMNZ structures, and have proposed a design using a single dielectric rod embedded in a metallic waveguide near its $TE_{10}$ cut-off frequency, imitating the medium with both effective relative permittivity and permeability near zero. We have numerically shown that this design exhibits the relevant features of the EMNZ structures.

Authors thank Dr. Arthur R Davoyan for useful discussions. This work is supported in part by the US Air Force Office of Scientific Research (AFOSR) Multidisciplinary University Research Initiatives (MURI) on Quantum Metaphotonics & Metamaterials, Award No. FA9550-12-1-0488.